\documentclass[preprint,aps,nofootinbib,showpacs,amsfonts,epsf]{revtex4-2}

\input{epsf.tex}

\usepackage{amsmath}

\renewcommand{\a}{\alpha}

\newcommand{\be}{\begin{equation}}
\newcommand{\ee}{\end{equation}}
\newcommand{\bea}{\begin{eqnarray}}
\newcommand{\eea}{\end{eqnarray}}
\newcommand{\ba}{\begin{array}}
\newcommand{\ea}{\end{array}}
\def\J#1#2#3#4{{#1} {\bf #2}, #3 (#4)}
\def\PRD{Phys. Rev. D}
\def\PR{Phys. Rev.}
\def\PRL{Phys. Rev. Lett.}

\def\JMP{J. Math. Phys.}

\def\CQG{Class. Quantum Grav.}

\def\PLB{Phys. Lett. B}

\begin{document}
\draft
\title{Dyonic black holes: The theory of two electromagnetic potentials}

\author{C. J. Ram\'irez-Valdez,$^{\dagger\ddagger}$ H. Garc\'ia-Compe\'an,$^\dagger$ and V.~S.~Manko$^\dagger$}
\address{$^\dagger$Departamento de F\'\i sica, Centro de Investigaci\'on y
de Estudios Avanzados del IPN, A.P. 14-740, 07000 Ciudad de
M\'exico, Mexico\\$^\ddagger$Theoretical Particle Physics and
Cosmology Group, Department of Physics, King's College London,
University of London, Strand, London, WC2R 2LS, U.K.}

\begin{abstract}
In the present paper we argue that the dyonic black hole spacetimes
must be studied within the theory of two electromagnetic potentials,
and we use the dyonic Reissner-Nordstr\"om solution to demonstrate
that the field of the monopole magnetic charge is correctly
described by the $t$-component of the dual electromagnetic
potential. As a result, the Dirac string associated with the
$\varphi$-component of the usual electromagnetic 4-potential becomes
just a mathematical object, without any physical content, that
arises in some calculations when one employs unsymmetrical
representations of the electromagnetic field. We use three
different, though equivalent, forms of the electromagnetic
energy-momentum tensor to calculate the Komar mass of the
Reissner-Nordstr\"om black hole, and in one case the Dirac string is
linked to the magnetic charge, in another to the electric charge,
while the third, symmetrical case, is string-free.
\end{abstract}

\pacs{04.20.Jb, 04.70.Bw, 97.60.Lf}

\maketitle

\newpage

\section{Introduction}

The idea that magnetic charges are located ``at the end of an
unobservable string, which is the line along which the
electromagnetic potentials are singular'' belongs to Dirac
\cite{Dir}, and one may think that it introduces certain physical
asymmetry between electricity and magnetism. This unsymmetrical
approach to electromagnetism was criticized by Schwinger \cite{Sch}
who advocated the symmetrical viewpoint embodying invariance under
charge rotation which leads to the integer quantization condition,
as opposed to Dirac's ``half-integer'' condition. An important
ingredient of Schwinger's symmetrical approach was the introduction
of a second electromagnetic vector potential defined nonlocally in
terms of the field strengths; it looks like Schwinger's remarkable
intuition was telling him that the magnetic charge cannot be
properly described by means of exclusively the ordinary potential
$A_\mu$.

The dyonic black hole solutions within the framework of general
relativity were first considered by Carter \cite{Car} who introduced
the magnetic charge parameter into the Reissner-Nordstr\"om (RN) and
Kerr-Newman \cite{KNe} spacetimes on physical grounds. However, his
analysis of the thermodynamic properties of black holes was
restricted to the case of zero magnetic charge only, most probably
to avoid the problem of singular electromagnetic sources. The dyonic
solutions widely arise also in other field theories (see, e.g.,
\cite{Cve1,Cve2}), which shows generic interest to the magnetic
monopoles in modern theoretical physics, thus motivating and
justifying efforts aimed at their correct description.

A few years ago, a discussion of the Dirac strings in dyons has
sprung up in relation to the problem of the mass distribution in the
dyonic KN black hole, when in the paper \cite{MGa} such distribution
was assumed to be the same as in the usual electrically charged KN
black hole, while in the paper \cite{CGa} a mathematical evaluation
of the mass integral gave rise to a model with two additional
semi-infinite massive sources due to Dirac strings. Although the
latter model was later criticized for its unphysical features
\cite{GMR}, we believe a convincing analytical demonstration of the
incorrectness of the entire Dirac-string concept is still needed to
clarify and broaden our knowledge about the dyonic spacetimes in
general and magnetic charges in particular. In the present paper,
the first of a short series of two papers, we consider the static RN
dyonic black hole solution which, in our opinion, is the best
example of a spacetime for the presentation and illustration of both
the basic ideas on the description of magnetic monopoles and the
related mathematical calculations, while in the second paper we
shall extend our approach to the stationary spacetimes and the
dyonic KN black hole. It is precisely the spherical symmetry of the
dyonic RN solution that helped us actually realize that the
$\varphi$-component of the potential $A_\mu$ is nothing more but an
auxiliary mathematical function whose singularity structure should
not be ascribed to the RN dyon itself, whereas the field of the
magnetic charge is correctly described by the $t$-component of the
dual electromagnetic potential $B_\mu$ that does share the spherical
symmetry of the RN spacetime. The reader will see that the presence
or absence of the string term in the mass integrals essentially
depends on the choice of the specific representation of the
energy-momentum tensor, and one representation even gives rise to a
``Dirac string'' associated with the electric charge.

Our paper is organized as follows. In the next section we consider
the Maxwell equations in the symmetrical form and give three
different, though equivalent, representations of the energy-momentum
tensor of the electromagnetic field in terms of the usual and dual
electromagnetic tensors. Here we also present the dyonic RN solution
and calculate two nonzero components of the corresponding dual
4-potential $B_\mu$. In Sec.~III the Komar mass \cite{Kom} of the
dyon RN solution is calculated in three different ways, clearly
demonstrating the auxiliary mathematical character of the components
endowed with singular Dirac strings. The results obtained are
discussed in Sec.~IV.

\section{Two-potential formulation of Maxwell's equations and the dyonic RN solution}

Motivated by Schwinger's symmetrical approach to the description of
dyons \cite{Sch}, we write the vacuum Maxwell equations in the
absence of currents in the form
\be \partial_\nu(\sqrt{-g}F^{\mu\nu})=0, \quad
\partial_\nu(\sqrt{-g}\tilde F^{\mu\nu})=0, \label{ME} \ee
where
\be \tilde F^{\mu\nu}
=\frac{1}{2}\epsilon^{\mu\nu\a\beta}F_{\a\beta} \label{tF} \ee
is the dual electromagnetic tensor.

Eqs. (\ref{ME}) imply the existence of the potentials $A_\nu$ and
$B_\nu$, such that
\be F_{\mu\nu}=\partial_\mu A_\nu-\partial_\nu A_\mu, \quad \tilde
F_{\mu\nu}=\partial_\mu B_\nu-\partial_\nu B_\mu,\label{AAt} \ee
which in the language of differential forms rewrites as
\be F=dA, \quad \star F=dB, \label{fA} \ee
the star symbol denoting Hodge dual.

The energy-momentum tensor of the electromagnetic field is normally
taken in the form
\be
T{^\mu}{_\nu}=\frac{1}{4\pi}\left(F^{\mu\alpha}F_{\nu\alpha}-\frac{1}{4}\delta^\mu_\nu
F^{\alpha\beta}F_{\alpha\beta}\right), \label{TE1} \ee
and, as will be shown in the next section, it is precisely this
representation of $T{^\mu}{_\nu}$ that leads to appearance of
singular terms due to magnetic field in the mass integrals. Apart
from (\ref{TE1}), it is advantageous to have two other equivalent
representations of $T{^\mu}{_\nu}$ which involve the dual
electromagnetic tensor $\tilde F^{\mu\nu}$. For this purpose we use
the identity \cite{Whe}
\be
A^{\mu\alpha}B_{\nu\alpha}-\tilde{A}^{\mu\alpha}\tilde{B}_{\nu\alpha}=\frac{1}{2}\delta^\mu_\nu
A^{\alpha\beta}B_{\alpha\beta}, \label{Id} \ee
which is valid for any two antisymmetric tensors $A^{\mu\nu}$ and
$B_{\mu\nu}$, and their duals $\tilde A^{\mu\nu}$ and $\tilde
B_{\mu\nu}$. Then the second representation of $T{^\mu}{_\nu}$ takes
the symmetrical form
\be T{^\mu}{_\nu}=\frac{1}{8\pi}(F^{\mu\alpha}F_{\nu\alpha}
+\tilde{F}^{\mu\alpha}\tilde{F}_{\nu\alpha}), \label{TE2} \ee
while for the third representation in terms of the dual tensor only
we get
\be T{^\mu}{_\nu}=\frac{1}{4\pi}\left(\tilde F^{\mu\alpha}\tilde
F_{\nu\alpha}-\frac{1}{4}\delta^\mu_\nu \tilde F^{\alpha\beta}\tilde
F_{\alpha\beta}\right). \label{TE3} \ee

It is our purpose to demonstrate that the field of the magnetic
charge is better described by the dual potential $B_\mu$ than by
$A_\mu$. So, we can take a dyonic RN black hole as the simplest
model for our analysis, described by the metric \cite{Car}
\be ds^2=-f dt^2+f^{-1} dr^2+r^2(d\theta^2+\sin^2\theta d\varphi^2),
\quad f=1-\frac{2m}{r}+\frac{q^2+p^2}{r^2}, \label{RNm} \ee
with the corresponding electromagnetic field defined by the 1-form
\be A=A_tdt+A_\varphi d\varphi=-\frac{q}{r}dt-p\cos\theta d\varphi,
\label{RNA} \ee
where $m$, $q$ and $p$ are the parameters of mass, electric charge
and magnetic charge, respectively.

The RN metric (\ref{RNm}) represents a static spherically symmetric
spacetime of point charges for which we now should calculate the
components of the dual potential $B_\nu$. These can be found by
solving the following differential equations:
\be \partial_r B_t=-\frac{1}{r^2\sin\theta}\partial_\theta
A_\varphi, \quad \partial_\theta B_\varphi =r^2\sin\theta\partial_r
A_t, \label{sys} \ee
which are obtainable from the second equation in (\ref{fA}) by
taking the dual of $F$ and by noting that $dB=d(B_\nu dx^\nu)$. From
(\ref{RNA}) and (\ref{sys}) we readily get
\be B=B_tdt+B_\varphi d\varphi=\frac{p}{r}dt-q\cos\theta d\varphi,
\label{At} \ee
where the integration constants have been assigned zero values.

By comparing the expressions (\ref{RNA}) and (\ref{At}), we can see
that both $A$ and $B$ have a well-behaved $t$-component, as well as
a string $\varphi$-component. Taking into account the spherical
symmetry of the dyonic RN spacetime, it would be plausible to draw a
conclusion that the electric field is determined by the
$t$-component $A_t=-q/r$ of $A$, whereas the magnetic field is
defined by the $t$-component $B_t=p/r$ of $B$, both $A_t$ and $B_t$
sharing spherical symmetry of the RN solution. In this respect,
having two electromagnetic potentials at hand, the affirmation that
the magnetic monopole charge $p$ is described by the string
component $A_\varphi=-p\cos\theta$ would be equivalent to affirming
that the electric field of a point-like charge $q$ is defined by
$B_\varphi=-q\cos\theta$, with an ``electric Dirac string''
consisting of two semi-infinite singularities at $\theta=0,\pi$.
Therefore, in view of the auxiliary mathematical role of the
components $A_\varphi$ and $B_\varphi$ it would be obviously wrong
to ascribe the string singularity of the former to the field of the
magnetic charge $p$, and the string singularity of the latter to the
field of the electric charge $q$, on equal grounds.

We shall now illustrate a purely mathematical character of the
components $A_\varphi$ and $B_\varphi$ by calculating the Komar mass
of the dyonic RN solution in three different ways.

\section{Calculation of the Komar mass integral}

The Komar mass is defined by the surface integral
\be M_K=-\frac{1}{8\pi}\int_\infty\star dk, \label{Km} \ee
where $k=g_{tt}dt$ is the covector associated to the timelike
Killing vector $\partial_t$.

Let us first see how (\ref{Km}) can be evaluated straightforwardly
just using the metric (\ref{RNm}), for which purpose we calculate
(\ref{Km}) for some sphere of constant radius $r$ and then take the
limit $r\to\infty$. By noting that in our case
\be \star dk=\partial_r g_{tt}r^2\sin\theta d\theta\wedge d\varphi,
\label{Intd} \ee
we have
\be M_r=-\frac{1}{8\pi}\int_{r=\rm const}(-\partial_r f)
r^2\sin\theta d\theta d\varphi=\frac{1}{2}r^2\partial_r f
=m-\frac{q^2+p^2}{r}, \label{Mr} \ee
so that
\be M_K=\lim_{r\to\infty}M_r=m. \label{Mm} \ee

To analyze the contribution of the electromagnetic field into the
mass integral (\ref{Km}) in more detail, it is advantageous to
rewrite (\ref{Km}) in the form
\be M_K=\frac{1}{4\pi}\int_{\infty} D^\nu k^\mu
d\Sigma_{\mu\nu}=\frac{1}{4\pi}\int_{\partial \mathcal{M}} D^\nu
k^\mu d\Sigma_{\mu\nu} +\frac{1}{4\pi}\int_{\mathcal{M}}D_\nu D^\nu
k^\mu dS_\mu \label{KmO} \ee
by means of Ostrogradsky's formula, where $k^\mu=\delta_t^\mu$. If
$\partial \mathcal{M}$ is chosen as a sphere of constant radius $r$,
then the first integral on the right-hand side of (\ref{KmO}) is
just $M_r$ in (\ref{Mr}), and in particular if
$r=r_+=m+\sqrt{m^2-q^2-p^2}$, $r_+$ being the radius of the event
horizon (the case that is of interest to us), then
\be \frac{1}{4\pi}\int_{H} D^\nu k^\mu
d\Sigma_{\mu\nu}=m-\frac{q^2+p^2}{r_+}. \label{KmH} \ee

Following \cite{CGa}, we now introduce the electromagnetic field
explicitly into the ``geometrical'' formula for $M_K$ by writing the
bulk integral from (\ref{KmO}) in the form
\be \frac{1}{4\pi}\int_{\mathcal{M}}D_\nu D^\nu k^\mu dS_\mu
=-2\int_{\mathcal{M}} T{^\mu}{_\nu} k^\nu dS_\mu \label{RT} \ee
with the aid of the well-known relations
\be D_\nu D^\nu k^\mu=-R{^\mu}{_\nu} k^\nu=-8\pi T{^\mu}{_\nu}k^\nu.
\label{kRT} \ee

Below we will calculate the integral on the right-hand side of
(\ref{RT}) for three different (but equivalent) representations
(\ref{TE1}), (\ref{TE2}) and (\ref{TE3}) of the energy-momentum
tensor $T{^\mu}{_\nu}$. Of course, in all three cases we must get
the same result $(q^2+p^2)/r_+$, as the integrals (\ref{Mm}) and
(\ref{KmH}) are known.

\subsection{The canonical representation (\ref{TE1})}

Note that in this representation the bulk integral (\ref{RT}) will
contain the function $A_\varphi$ explicitly after the Ostrogradsky
theorem  is applied for converting (\ref{RT}) into the surface
integral, and hence the contribution of the ``magnetic Dirac
string'' must be taken into account. Bearing this in mind, we get
\bea \frac{1}{4\pi}\int_\mathcal{M}D_\nu D^\nu k^\mu dS_\mu
&=&-2\int_\mathcal{M} T{^t}{_t}\sqrt{-g}\,d^3x \nonumber\\
&=&-\frac{1}{4\pi}\int_\mathcal{M}(F^{ta}F_{ta}
-F^{\varphi a}F_{\varphi a})\sqrt{-g}\,d^3x \nonumber\\
&=&\frac{1}{4\pi}\int_\mathcal{M}\partial_a
[\sqrt{-g}(F^{ta}A_t-F^{\varphi a}
A_\varphi)]d^3x \nonumber\\
&=&\frac{1}{4\pi}\int_{\Sigma_a}
(F^{ta}A_t-F^{\varphi a}A_\varphi)d\Sigma_a \nonumber\\
&=&\frac{1}{4\pi}\int_H F^{tr} A_t\,d\Sigma_r-\frac{1}{4\pi}\int_S
F^{\varphi\theta} A_\varphi\,d\Sigma_\theta, \label{BI1} \eea
where `$H$' refers to the horizon and `$S$' refers to the string. In
the last step we have taken into account that
\be \int_S F^{t\theta} A_t\,d\Sigma_\theta=0, \quad \int_H
F^{\varphi r} A_\varphi\,d\Sigma_r=0, \label{zi1} \ee
because $F^{t\theta}=0$ and $F^{\varphi r}=0$. Finally, we readily
obtain
\bea &&\int_H F^{tr} A_t\,d\Sigma_r =\int_H F_{tr} A_t r^2\sin\theta
d\theta d\varphi=4\pi q^2/r_+, \nonumber\\ &&\int_S
F^{\varphi\theta} A_\varphi\,d\Sigma_\theta
=2\lim_{\theta\to\pi}\int_{r_+}^\infty\int_0^{2\pi}
F_{\varphi\theta} A_\varphi\frac{1}{r^2\sin\theta}dr d\varphi=-4\pi
p^2/r_+, \label{HS1} \eea
which leads to $(q^2+p^2)/r_+$ for (\ref{BI1}).

Note that in this representation of $T{^\mu}{_\nu}$ the contribution
of the electric charge into the bulk integral (\ref{BI1}) comes from
the horizon, and the contribution of the magnetic charge comes from
the string.

\subsection{The dual representation (\ref{TE3})}

This case is fully analogous to the previous one, with the roles of
the electric and magnetic fields interchanged:
\bea  \frac{1}{4\pi}\int_{\mathcal{M}}D_\nu D^\nu k^\mu dS_\mu
&=&-\frac{1}{4\pi}\int_\mathcal{M} (\tilde{F}^{ta}
\tilde{F}_{ta}-\tilde{F}^{\varphi a}\tilde{F}_{\varphi a})
\sqrt{-g}\,d^3x \nonumber\\
&=&\frac{1}{4\pi}\int_\mathcal{M} (\tilde{F}^{ta}\partial_a
B_t-\tilde{F}^{\varphi a}\partial_a
B_\varphi) \sqrt{-g}\,d^3x \nonumber\\
&=&\frac{1}{4\pi}\int_\mathcal{M} \partial_a [\sqrt{-g}
(\tilde{F}^{ta}B_t-\tilde{F}^{\varphi a}
B_\varphi)] d^3x \nonumber\\
&=&\frac{1}{4\pi}\int_{\Sigma_a}(\tilde{F}^{ta}
B_t-\tilde{F}^{\varphi a}B_\varphi)d\Sigma_a \nonumber\\
&=&\frac{1}{4\pi}\int_H \tilde{F}^{tr} B_t\,d\Sigma_r
-\frac{1}{4\pi}\int_S \tilde{F}^{\varphi\theta}
B_\varphi\,d\Sigma_\theta, \label{BI2} \eea
where we have taken into account that
\be \int_S \tilde{F}^{t\theta} B_t \,d \Sigma_\theta=0
\quad\mbox{and}\quad \int_H \tilde{F}^{\varphi r} B_\varphi \,
d\Sigma_r=0. \label{zi2} \ee

The evaluation of the last two integrals in (\ref{BI2}) yields
\be \int_H \tilde{F}^{tr} B_t \,d \Sigma_r=4\pi p^2/r_+, \quad
\int_S \tilde{F}^{\varphi\theta} B_\varphi \, d\Sigma_\theta=-4\pi
q^2/r_+, \label{int2} \ee
and in this representation of the energy-momentum tensor it is the
electric charge that develops an ``electric Dirac string'', so that
this time the electrostatic contribution into the bulk integral
(\ref{BI2}) comes from the string, while the contribution of the
magnetic charge comes from the horizon!

\subsection{The symmetrical representation (\ref{TE2})}

In this representation only the well-behaved components of the
electromagnetic potentials are involved in the calculations of the
bulk integral (\ref{RT}), so that no any auxiliary string
contribution arises during the application of Ostrogradsky's theorem
converting the bulk integral into the surface integral:
\bea \frac{1}{4\pi}\int_{\mathcal{M}}D_\nu D^\nu k^\mu
dS_\mu&=&-\frac{1}{4\pi}\int_{\mathcal{M}}
(F^{ta}F_{ta}+\tilde{F}^{ta}\tilde{F}_{ta})\sqrt{-g}\,d^3x \nonumber\\
&=&\frac{1}{4\pi}\int_{\mathcal{M}} (F^{ta}\partial_a
A_t+\tilde{F}^{ta}
\partial_a B_{t})\sqrt{-g}\,d^3x \nonumber\\
&=&\frac{1}{4\pi}\int_{\mathcal{M}}\partial_a [\sqrt{-g}
(F^{ta}A_t+\tilde{F}^{ta}B_{t})]\,d^3x \nonumber\\
&=&\frac{1}{4\pi}\int_{\Sigma_a} (F^{ta}A_t
+\tilde{F}^{ta}B_{t}) d\Sigma_a \nonumber\\
&=&\frac{1}{4\pi}\int_H F^{tr} A_t \, d\Sigma_r+\frac{1}{4\pi}\int_H
\tilde{F}^{tr} B_t \, d\Sigma_r \label{BI3}, \eea
and evaluation of the last two integrals readily gives
\be \int_H F^{tr} A_t \, d\Sigma_r=4\pi q^2/r_+, \quad \int_H
\tilde{F}^{tr} B_t \, d\Sigma_r=4\pi p^2/r_+. \label{int3} \ee

Therefore, in the symmetrical representation of $T{^\mu}{_\nu}$, the
calculation of the Komar mass of the dyonic RN source reduces to
evaluation of the surface integrals over the event horizon only. As
we have shown, the choice of the particular representation does not
alter the final result when the singularity structure of the
functions involved in the concrete calculational scheme is carefully
taken into account.

\section{Discussion and conclusions}

The analysis carried out in the previous two sections clearly shows
that the problem of the Dirac string associated in the literature
with the magnetic charge is actually an artificial mathematical
issue arising as a result of a wrong identification of the potential
describing the field of the magnetic monopole. Thus we have seen
that the same contributions into the mass integral can be made by
the horizon or string terms, and these are interrelated as follows:
\bea  &&\int_H F^{tr} A_t \, d\Sigma_r=-\int_S
\tilde{F}^{\varphi\theta}
B_\varphi d\Sigma_\theta=4\pi q \Phi_e, \nonumber\\
&&\int_H \tilde{F}^{tr} B_t \, d\Sigma_r=-\int_S F^{\varphi\theta}
A_\varphi \, d\Sigma_\theta=4\pi p \Phi_m, \label{isum} \eea
where we have introduced the horizon values of the electric and
magnetic potentials $\Phi_e$ and $\Phi_m$ by the well-known formulas
\be \Phi_e=q/r_+, \quad \Phi_m=p/r_+, \label{Fem} \ee
and now it is manifest that $\Phi_m$ is just the dual component
$B_t$ evaluated on the horizon.

It should be also stressed that the distribution of the Komar mass
along the horizon and the magnetic (or electric) string singularity
of the component $A_\varphi$ (or $B_\varphi$) appearing during the
computation of the mass integral (\ref{Km}) is just a mathematical
abstraction that should not be interpreted as reflecting the real
physical distribution of mass in the dyonic RN black hole, which is
of course spherically symmetric. In this respect it would probably
be worth drawing analogy with the static vacuum Weyl gravitational
fields which all satisfy the Laplace equation $\Delta\psi=0$ for an
auxiliary function $\psi$, but the real physical field is
$f=\exp\psi$ which apparently has a different singularity structure
than $\psi$.

A curious feature of the bulk integral (\ref{RT}) additionally
pointing at its auxiliary technical character is that it does not
seem to be actually involved in the Smarr mass formula \cite{Sma},
the latter important relation following directly from the surface
integral (\ref{KmH}) evaluated on the horizon. Indeed, after
rewriting (\ref{KmH}), on  the one hand, in terms of the potentials
$\Phi_e$ and $\Phi_m$ as
\be \frac{1}{4\pi}\int_{H} D^\nu k^\mu
d\Sigma_{\mu\nu}=m-q\Phi_e-p\Phi_m, \label{KmH2} \ee
and recalling, on the other hand, that, as was shown by Carter
\cite{Car},
\be \frac{1}{4\pi}\int_{H} D^\nu k^\mu
d\Sigma_{\mu\nu}=\frac{\kappa}{4\pi}\mathcal{A}, \label{KmH3}  \ee
where $\kappa$ is the surface gravity and $\mathcal{A}$ the area of
the event horizon, we immediately arrive at the Smarr relation
verified by the dyonic RN black hole:
\be m=\frac{\kappa}{4\pi}\mathcal{A}+q\Phi_e+p\Phi_m. \label{RNS}
\ee

As a final remark, it would be probably worth mentioning that our
results suggesting the non-existence of magnetic and electric Dirac
strings are particularly important in application to the systems of
many dyonic black holes, for which a correct calculation of
individual Komar masses would be practically impossible in the
presence of numerous string singularities. Our symmetrical approach
in which the individual masses are evaluated on the horizons, and
hence are entirely located inside the horizons, does not have this
kind of problem, confirming for instance the definition of the Komar
mass in a binary system of magnetically charged Reissner-Nordstr\"om
black holes \cite{MRS}.

\section*{Acknowledgments}

This work was partially supported by CONACyT of Mexico and by
``Secretar\'ia de Educaci\'on, Ciencia, Tecnolog\'ia e Innovaci\'on
de la Ciudad de M\'exico (SECTEI)'' of the Mexico City.

\end{document}